# SoMIAP: Social media images analysis and prediction framework

*Yonghao Shi[1], Gueltoum Bendiab[2], Stavros Shiaeles[2], and Nick Savage[2]*

*CSCAN, University of Plymouth, Plymouth, UK, PL4 8AA*

*syh506795003@yahoo.com*

*Cyber Security Research Group, University of Portsmouth, UK*

*gueltoum.bendiab@port.ac.uk, sshiaeles@ieee.org, nick.savage@port.ac.uk*

**Abstract**. The personal photos captured and submitted by users on social networks can provide several interesting insights about the user's location, which is a key indicator of their daily activities. This information is invaluable for security organisations, especially for security monitoring and tracking criminal activities. Hence, we propose in this paper a novel approach for location prediction based on the image analysis of the photos posted on social media. Our approach combines two main methods to perform the image analysis: place and face recognition. The first method is used to determine the location area in the analysed image. The second is used to identify people in the analysed image, by locating a face in the image and comparing it with a dataset of images that have been collected from different social platforms. The effectiveness of the proposed approach is demonstrated through performance analysis and experimental results.

**Keywords:** Image analysis · place recognition · face recognition · location prediction · security · social media.

## I. Introduction

The large number of personal photos shared on social media platforms presents a new opportunity to develop location prediction systems [27]. In this context, Instagram is a popular social network for sharing real-time photos with over95 million photos uploaded each day [17]. These photos commonly contain sensitive information about the user, like places they usually go to, whether or not they are on vacation, and who are their friends and family members [28, 27]. Thus, hundreds of millions of shared photos can provide several interesting insights on the locations of people [9]. This can lead to different use cases of such information, as it is a key indicator of their daily activities. Image analysis to predict and identify locations on social media could produce effective real-time monitoring systems that can be relevant to law enforcement authorities and others involved with public security to fight criminality and terrorism [13].

It may also keep track of malicious actions and threats such as the paedophile hunter activities grooming and planned criminal activities; where a person's current location could be coupled with geocoded crime statistics to predict when and where crimes will occur [13]. In this context, a recent report of the Guardian stated that complaints to police about alleged crimes linked to the use of Facebook and Twitter have increased by 780% in four years. While, in 2018, there were more than 4,908 crime reports in the UK, which involved the two sites [2].Moreover, in the case of an emergency, incident or crisis, local authorities can achieve situational awareness in a short space of time because of the speed of social media in providing visual snapshots of the incidents that took place [9].

Motivated by the above use-cases and others, a wide array of approaches has been developed to extract the location information of users on social media [8, 23]. A small number of researchers have previously exploited images submitted on those platforms to extract such information e.g. [28], by focusing on non- visual features of social media images such as geotags, descriptions, and other metadata [28, 9]. Most of these approaches have limitations regarding coverage because a large number of the submitted photos may contain poor annotation, or no annotation at all [28], which affects their accuracy and raises questions about their effectiveness in real-world scenarios. In this paper, we aim to address these issues, by proposing a novel approach for predicting a user's locations through submitted photos on online social media. Our approach combines two main image analysis processes: place location and face recognition. The place location process is used to find the location area in the analysed image, by comparing it with a data set of known location images. While the face recognition technology is used to detect people in the analysed image by locating their face in the image and comparing it with a dataset of images that have been collected from different social platforms. The combination of those two techniques has the benefit of providing high accuracy and potentially complete coverage needed for security monitoring.

The rest of the paper is organized as follows. Section II presents the related works on the topic of identifying a user's location through social media. Section III provides a detailed explanation of the proposed approach. The experimental results and discussions are presented in Section IV. Finally, Section V reviews the content of the paper, presents the conclusions, and outlines the potential future work.

## II. Related work

The prediction of a user's geographical location through social media has attracted intensive attention in recent years. In this context, a variety of information sources have been exploited for predicting locations. Some researchers have exploited a user's relationships on social networks to predict their location, such as works in [19, 10, 21]. These approaches are built upon the finding that friends in social media will be very likely to live in the same location. Therefore, the location of a given user can be predicted based on their relationship with other users. Most of these social graph-based approaches investigated inductive machine learning methods for making the predictions. Thus, the accuracy of the proposed approach highly depends on the samples used to train the classifier.

Another kind of approach used the content associated with a user's posts to estimate their locations [8, 12, 6, 5, 23]. These approaches generally predict a user's locations by examining words from their profile and generated posts. For example, authors in [8] used a user's tweets to extract words that are highly related to a specific geographic region and used these words to calculate the prob- ability that this user lives at a specific location. The user profile is also applied in several content-based approaches to establish the location of users, particularly in twitter [19, 3, 15], where user profile is often combined with other information from message metadata such as MML (latitude and longitude geotags) [15], or other services such as Google Maps use geocoding to generate coordinates from a place name [3]. Most of these approaches face several key challenges, such as the uncertainty and heterogeneity of the generated content, which can affect the accuracy of the estimator. Study in [32] affirmed that content-based approaches have moderate coverage and low accuracy because of the lack of relationship between the user's true physical location and the mentioned one in the posts, as many users provide invalid place names. For example, on Twitter, from 46% to 77% of users provide fake location information [15].

Although a limited number of studies have focused on shared images through social media to predict a user's location, some investigative effort has been made to exploit the non-visual features of social media images such as geotags, descriptions, and other attributes to determine locations. For instance, work in [28] investigated the fusibility of geotagged photos of Flickr to examine the spatiotemporal human activities, which can be used for further location analysis. This work highlighted the potential of Volunteered Geographic Information (VGI) to examine human activities in space and time. In other recent work [9], authors introduced an image classification model to detect geotagged photos that are further analysed to determine if a fire event did occur at a particular time and place. To evaluate the proposed approach, the authors used a dataset of 114,098 Flickr photos, where 20.3% of the collected photos were geotagged and 2.5% contained GPS data in the EXIF (Exchangeable image file) header. The authors reported an average accuracy of 132 km in identifying the location of fires. However, those approaches have low coverage as only 2.5% of flicker photos have EXIF header information [27].

Furthermore, data is often noisy, and photos may contain poor annotation, or no annotation at all, which affect the accuracy of these approaches. In order to overcome the drawbacks of non-visual features-based approaches, some researchers have exploited the visual features of images to extract the location information. For example, work in [7] proposed visual features that assess the images at a low level. In this work, the authors performed image analysis of photos posted on Twitter by extracting SIFT (Scale-Invariant Feature Transform) descriptors [18] from the images. Then, the SIFT descriptors were clustered to form visual words. Authors reported an average F1-score of 70.5% of image classification using text, image, and social context features. The main problem of this work is the lack of geotagged images. In prior work [14], authors proposed a data-driven scene matching approach to predict the location of social geotagged images at the global scale, they first retrieve visually similar photos and form clusters using geo-clustering. The geo-centroid of the cluster containing the most photos used for location prediction.

## III. Proposed approach

The main idea of this study is to present a visual-content-based approach that predicts a user's location from photos on social media. Our methodology comprises of two main image analysis steps. Firstly, we apply a place recognition technique on the input image to find the location area where the image was taken. This process is done by matching the input image with a data set of known-location images. In the second step, a face recognition technique is used to detect people in the input image, by locating their face in the image and comparing it with a dataset of images that have been collected from different social platforms. By combining the results of the two steps, the system outputs the location where the image is taken with the name of the person or persons shown in the image. More details are given in the following sections.

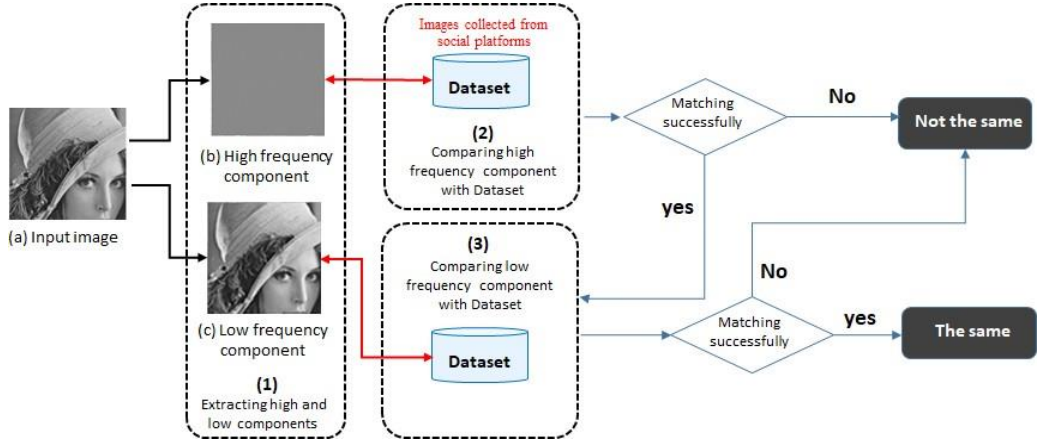

**Fig. 1**: Overview of the place location process.

### A.   Place location

The location recognition step compares the input image with images of well- known locations to detect the area where the image was taken. As illustrated in Fig. 1, the place recognition process is performed by using the low and high frequency components, which are extracted from the input image. The low- frequency component provides details that describe the basic information of the input image, while the high-frequency component is very useful to find descriptive and invariant features for image matching. Firstly, the high-frequency component is compared with a dataset that contains samples of known locations images (*see* Fig. 1). If the features matching in the first step is successful, the low-frequency component is then used to compare the basic information of the input image with the samples in the dataset. In this step, the comparison is done by using an image hashing algorithm, which is a fast method to compare the differences between the images.

### B.   Face recognition

The face recognition process is used to identify if a person is in the input image, where his or her face is located, and who this person is. It has as input, the image to be analysed, and as output, the identity of the person that appears in this image. The procedure is separated into two main steps: face detection and face recognition.

**Face detection:** in this step, a Haar classifier [31] is used to determine the existence of human faces in the input image and their locations. This classifier uses a set of Haar features [31] to capture important characteristics of human faces in the input image, and removes all the unwanted objects (e.g., building, tree, etc.). Several research works have proven the effectiveness of Haar features in building accurate classifiers with a limited number of images in training the data set [31]. The expected outputs of this step are detected faces in the input image, which will be used for further processing.

**Face recognition:** the goal of this step is to automatically determine the identities of the faces in the input images. Thus, a faces dataset is required, where for each person, several images are taken, and their features are extracted and stored in the dataset. In our approach, the facial images were collected from different social platforms. Then, features are extracted from each input facial image and compared with those of each face class stored in the database. For the comparison, the Fisherface method [29] is used when the facial images have large variations in illumination and facial expression. Otherwise, the Local Bi- nary Patterns Histograms (LBPH) [1] algorithm is more suitable. Fig. 2 gives an overview of the face recognition flowchart.

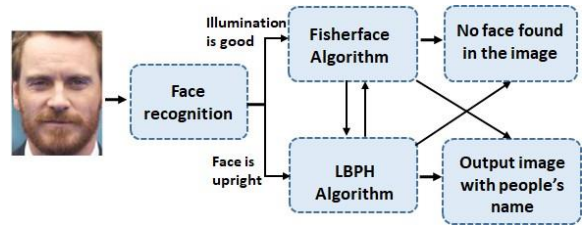

**Fig. 2.** Face recognition flowchart.

### IV.   Experiments and Discussion

In this section we describe the experiments and results obtained from the methods shown in section III. First, we describe the experimental set-up and results for each step. Then we compare the performance of the proposed method with baseline methods.

### A.   Location recognition experiment results

In these experiments, we used the OpenCV (Open-

Source Computer Vision) library [20], where the three descriptors; SIFT, Speeded Up Robust Features (SURF) [4], and Oriented FAST and Rotated BRIEF (ORB) [24] were used for feature detection and collection. The hashing algorithms dhash [11] and pHash [22] were used to compare the differences between images; grayscale-images and colour images. For the training and testing, we used a dataset of 400 well-known location photos collected from social platforms: Facebook, Wechat and Tecent, with 200 similar images and 200 different images.

**High-frequency component:** Fig 3. shows comparison results for the SIFT, SURF and ORB techniques. The results illustrate the robustness and accuracy of the SURF algorithm for feature detection. While the ORB algorithm is more suitable to solve the rotation problem of image (Fig. 4). ORB provides faster image matching than the SURF, but it is less accurate.

**Low-frequency component:** For the low-frequency component experiments, four tests were carried out to determine the accuracy of the dhash and pHash algorithms. Tests were performed between similar images and different images, by using a dataset consisting of 200 similar images and 200 different images. For each test, we identified the weight hash values for the similar images test and the different images test, the threshold, and the accuracy. Two tests were performed with grayscale images and the two others with colour-images. In the grayscale image tests, the hash value is computed after converting the image to the grayscale format. In the colour image tests, the hash values were calculated for each RGB colour separately. After calculated the hash value for Blue is appended to Green and finally both to Red.

As shown in the Table 1, the dhash algorithm achieved the highest accuracy with colour images (64.4%) for a threshold of 34. Whereas the pHash algorithm reached an accuracy of 64% for the grayscale-images test with a threshold of 23, which is higher than the dHash one with approximately 7 percentage points.

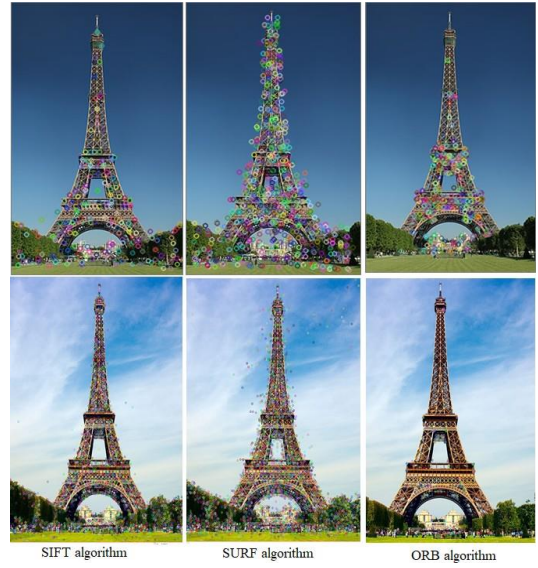

**Fig.3.** Comparison of the SIFT, SURF and ORB algorithms.

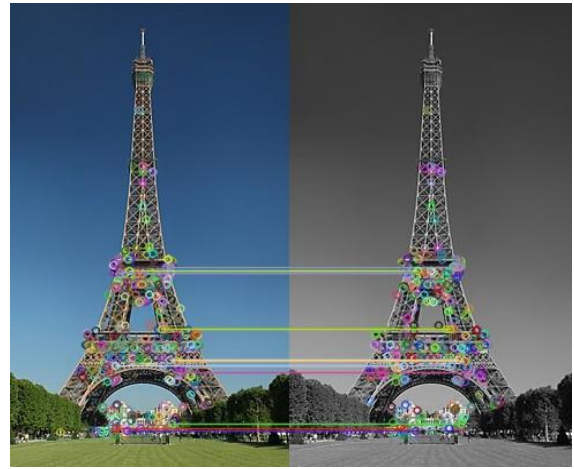

**Fig. 5.** Results for illumination.

Similarly, it achieved the highest accuracy for the colour images test (68%) for the same threshold (23). From the results, we concluded that the pHash algorithm has higher accuracy than the dhash for grayscale-images and colour images. Also, the hashing algorithms pHash and dhash provide higher accuracy with colour images than grayscale images.

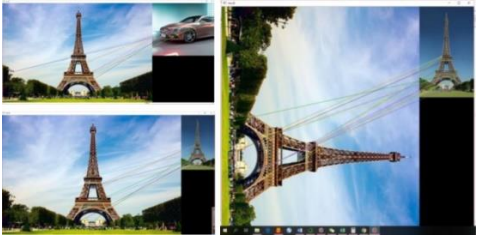

**Fig. 4.** Results for rotation.

**Table 1.** Tests results for Dhash and PHash algorithm

| Tests | Weight value | | Threshold | Accuracy |
|---|---|---|---|---|
| | Similar images | Different images | | |
| dhash (grayscale) | 30.35 | 40.37 | 34 | 59.6% |
| dhash (colour) | 32.67 | 40.66 | 36 | 64.4% |
| pHash (grayscale) | 18.43 | 27.81 | 23 | 64.0% |
| pHash (colour) | 19.83 | 23.00 | 23 | 68.0% |

Table 2 shows the min, max and average times spend on data analysis for each test. The results reveal that for the two algorithms, the data analysis take more time for colour images than the grayscale images. Also, the pHash is a time-consuming algorithm (maximum time cost is 1.25s) compared to the dhash (maximum time cost is 0.6s). In summary, pHash algorithm has higher accuracy, but lower speed. While the dhash algorithm is fast but less accurate. In fact, pHash has the highest accuracy in all hash algorithms. However, it is time-consuming because of the complex calculations. Also, the two hashing algorithms provide higher accuracy with colour images but they need more computing time because the hash value calculation time is tripled.

**Table 2.** Speed tests results for dhash and PHash algorithm

| Tests | Mean | Maximum | Minimum |
|---|---|---|---|
| Dhash (grayscale) | 0.10s | 0.50s | 0.10s |
| Dhash (colour) | 0.10s | 0.60s | 0.10s |
| PHash (grayscale) | 0.30s | 0.80s | 0.15s |
| PHash (colour) | 0.50s | 1.25s | 0.20s |

### B. Face recognition Experiments results

In the face recognition experiments, we used a Haar classifier to identify and collect front faces from the input image with rectangles (see Fig. 5). To recognize the collected front faces, we used the Fisherface, LBPH and Eigenface [30] algorithms. In these initial experiments, only 20 faces were used for testing and training. Fig.6 shows the images of three persons used for the training step.

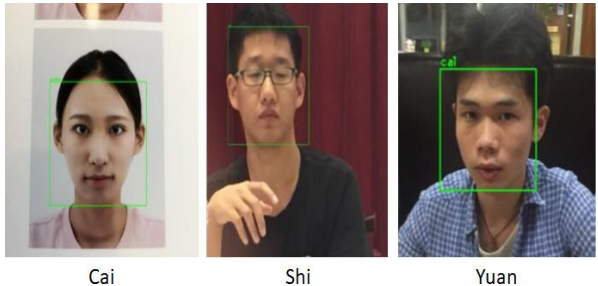

**Fig. 6.** Cai, Shi, and Yuan images used for the training and testing steps.

Table 3 shows the results for the face recognition using the three algorithms Fisherface, Eigenface and LBPH. From the results we can see that the Fisherface has the highest accuracy in face recognition, while the Eigenface algorithm is the faster. Fisherface is one of the popular algorithms used in face recognition and is widely believed to be superior to Eigenface. Thus, it is more suitable for our approach. However, the LBPH algorithm is more suitable in case of upright images.

**Table 3.** Eigenface, Fisherface and LBPH results

| Parameters | Eigenface | Fisherface | LBPH |
|---|---|---|---|
| Accuracy | 40% | 53.3% | 79% |
| Speed | Fast | Normal | Normal |
| Colour-element | Yes | Yes | Yes |
| Illumination | No | No | Yes |
| Rotation | Less than 10 | Less than 10 | No |

### C. Comparison

It is not easy to conduct a fair comparison among various image and location recognition approaches due to the differences between the datasets and algorithms used. Thus, our comparison will be based on some significant features including accuracy, speed, image type, rotation, and illumination. Table 4 overviews a general comparison between our approach and other approaches. Firstly, the hashing method is compared with other representative's approaches (*see* Table 4). Then, the proposed face recognition method is compared with previous works in the field (*see* Table 4). As can be seen from Table 4, the hashing method in our approach has the highest accuracy with a suitable threshold for colour images. Compared to the grayscale approaches, it increased by 5% and 4% for dhash and pHash algorithms respectively, which prove the effectiveness of computing the hash values from primary colours; red, blue and green. However, the results for the face recognition step need further investigation due to the limitation of training data.

### I. Conclusions

This paper proposed a novel approach for location prediction based on image analysis of the photos posted on social media. This approach exploited the visual features of images to extract the location information, by using the place and face recognition techniques. From our initial experimental results, the method seems promising and being able to predict the people locations, with potentially complete coverage required for security monitoring.

Future work would improve the accuracy of the face recognition process by using more samples for training and testing. Also, we intend to use other machine learning approaches such as to enhance the accuracy of this approach.

**Table 4.** Comparison with other methods

| Hashing method, comparison with other methods | | | | | |
|---|---|---|---|---|---|
| | Accuracy | Speed | Image type | Rotation | Illumination |
| Zauner et al [34] | 49.6% | 9.69s | Grayscale | Yes | Not mentioned |
| Yang et al [33] | 60% | 9.0.7s | Grayscale | Less than 30 | Yes |
| Jie et al [16] | 50% | Not mentioned | Grayscale | Less than 5 | Yes |
| Our approach | 66% | 0.5s | Colour | Yes | Yes |
| Face recognition method, comparison with other methods | | | | | |
| Nikolaos et al [25] | 40% | 0.5s | Grayscale | No | Yes |
| Magali et al [26] | 79% | 2s | Grayscale | No | No |
| Our approach | 53.3% | 1s | Colour | No | Yes |

## ACKNOWLEDGMENT

This project has received funding from the European Union's Horizon 2020 research and innovation programme under grant agreement no. 786698. The work reflects only the authors' view, and the Agency is not responsible for any use that may be made of the information it contains.

## References

1. Ahonen, T., Hadid, A., Pietikainen, M.: Face description with local binary patterns: Application to face recognition. IEEE Transactions on Pattern Analysis & Machine Intelligence **28**(12), 2037–2041 (2006).

2. Association, P.: social media-related crime reports up 780% in four years. https://bit.ly/2HPQIBS, accessed: 2019-07-19.

3. Baucom, E., Sanjari, A., Liu, X., Chen, M.: Mirroring the real world in social media: twitter, geolocation, and sentiment analysis. In: Proceedings of the 2013 international workshop on Mining unstructured big data using natural language processing. pp. 61–68. ACM (2013)

4. Bay, H., Ess, A., Tuytelaars, T., Van Gool, L.: Speeded-up robust features (surf). Computer vision and image understanding 110(3), 346–359 (2008)

5. Chandra, S., Khan, L., Muhaya, F.B.: Estimating twitter user location using social interactions–a content based approach. In: 2011 IEEE Third International Conference on Privacy, Security, Risk and Trust and 2011 IEEE Third International Conference on Social Computing. pp. 838–843. IEEE (2011)

6. Chang, H.w., Lee, D., Eltaher, M., Lee, J.: @ phillies tweeting from philly? predicting twitter user locations with spatial word usage. In: Proceedings of the 2012 International Conference on Advances in Social Networks Analysis and Mining (ASONAM 2012). pp. 111–118. IEEE Computer Society (2012)

7. Chen, T., Lu, D., Kan, M.Y., Cui, P.: Understanding and classifying image tweets. In: Proceedings of the 21st ACM international conference on Multimedia. pp. 781– 784. ACM (2013)

8. Cheng, Z., Caverlee, J., Lee, K.: You are where you tweet: a content-based approach to geo-locating twitter users. In Proceedings of the 19th ACM international conference on Information and knowledge management. pp. 759–768. ACM (2010)

9. Daly, S., Thom, J.A.: Mining and classifying image posts on social media to analyse fires. In: ISCRAM (2016)

10. Davis Jr, C.A., Pappa, G.L., de Oliveira, D.R.R., de L. Arcanjo, F.: Inferring the location of twitter messages based on user relationships. Transactions in GIS **15**(6), 735–751 (2011)

11. dhash: dhash 1.3. https://pypi.org/project/dhash/, accessed: 2019-07-19

12. Eisenstein, J., O'Connor, B., Smith, N.A., Xing, E.P.: A latent variable model for geographic lexical variation. In: Proceedings of the 2010 conference on empirical methods in natural language processing. pp. 1277–1287. Association for Computa- tional Linguistics (2010)

13. Hadjimatheou, K., Coaffee, J., De Vries, A.: Enhancing public security through the use of social media. European Law Enforcement Research Bulletin **18**, 1–14 (2019)

14. Hays, J., Efros, A.A.: Im2gps: estimating geographic information from a single image. In: 2008 ieee conference on computer vision and pattern recognition. pp. 1– 8. IEEE (2008)

15. Hecht, B., Hong, L., Suh, B., Chi, E.H.: Tweets from justin bieber's heart: the dynamics of the location field in user profiles. In: Proceedings of the SIGCHI conference on human factors in computing systems. pp. 237–246. ACM (2011)

16. Jie, Z.: A novel block-dct and pca based image perceptual hashing algorithm. arXiv preprint arXiv:1306.4079 (2013)

17. Lister, M.: 33 mind-boggling instagram stats facts for 2018. https://bit.ly/2zOKGfr, accessed: 2019-07-19

18. Lowe, D.G.: Distinctive image features from scale-invariant keypoints. Interna- tional journal of computer vision 60(2), 91–110 (2004)

19. McGee, J., Caverlee, J., Cheng, Z.: Location prediction in social media based on tie strength. In: Proceedings of the 22nd ACM international conference on Information & Knowledge Management. pp. 459–468. ACM (2013)

20. OpenCV: Opencv 4.1.1, the latest release is now

available. https://opencv.org/, accessed: 2019-07-19

21. Pellet, H., Shiaeles, S., Stavrou, S.: Localising social network users and profiling their movement. Computer Security 81, 49–57 (2018)

22. pHash: phash, the open-source perceptual hash library. https://www.phash.org/, accessed: 2019-07-19

23. Poulston, A., Stevenson, M., Bontcheva, K.: Hyperlocal home location identifica- tion of twitter profiles. In: Proceedings of the 28th ACM Conference on Hypertext and social media. pp. 45–54. ACM (2017)

24. Rublee, E., Rabaud, V., Konolige, K., Bradski, G.R.: Orb: An efficient alternative to sift or surf. In: ICCV. vol. 11, p. 2. Citeseer (2011)

25. Stekas, N., van den Heuvel, D.: Face recognition using local binary patterns his- tograms (lbph) on an fpga-based system on chip (soc). In: 2016 IEEE International Parallel and Distributed Processing Symposium Workshops (IPDPSW). pp. 300-304. IEEE (2016)

26. Stolrasky, M.S., Jakov, N.B.: Recognition using class specific linear projection. semanticscholar.org (2015)

27. Stone, Z., Zickler, T., Darrell, T.: Toward large-scale face recognition using social network context. Proceedings of the IEEE 98(8), 1408–1415 (2010)

28. Sun, Y., Fan, H., Helbich, M., Zipf, A.: Analyzing human activities through vol- unteered geographic information:

Using flickr to analyze spatial and temporal pat- tern of tourist accommodation. In: Progress in location-based services, pp. 57–69. Springer (2013)

29. Turk, M., Pentland, A.: Eigenfaces for recognition. Journal of cognitive neuro- science 3(1), 71–86 (1991)

30. Turk, M., Pentland, A.: Eigenfaces for recognition. Journal of cognitive neuro- science 3(1), 71–86 (1991)

31. Whitehill, J., Omlin, C.W.: Haar features for facs au recognition. In: 7th Inter- national Conference on Automatic Face and Gesture Recognition (FGR06). pp. 5–pp. IEEE (2006)

32. Xu, D., Cui, P., Zhu, W., Yang, S.: Find you from your friends: Graph-based residence location prediction for users in social media. In: 2014 IEEE international conference on multimedia and expo (ICME). pp. 1–6. IEEE (2014)

33. ang, B., Gu, F., Niu, X.: Block mean value-based image perceptual hashing. In: 2006 International Conference of Intelligent Information Hiding and Multimedia. pp. 167–172. IEEE (2006)

34. Zauner, C.: Implementation and benchmarking of perceptual image hash functions. http://phash.org/ (2010).